\documentclass[conference]{IEEEtran}
\IEEEoverridecommandlockouts
\usepackage{cite}
\usepackage{amsmath,amssymb,amsfonts}
 \usepackage{booktabs}
\usepackage{algorithmic}
\usepackage{graphicx}
\usepackage{textcomp}
\usepackage{xcolor}
\usepackage{hyperref}
\usepackage{graphicx}
\usepackage{caption}
\usepackage{subcaption}
\def\BibTeX{{\rm B\kern-.05em{\sc i\kern-.025em b}\kern-.08em
    T\kern-.1667em\lower.7ex\hbox{E}\kern-.125emX}}

\begin{document}
\title{Indian economy and Nighttime Lights\\
}
\author{\IEEEauthorblockN{Jeet Agnihotri}
\IEEEauthorblockA{\textit{ Computer Science and Engineering, M.B.M Engineering College } \\
\textit{ Jai Narain Vyas University}\\
Jodhpur, India \\
jeet99agnihotri.ju@gmail.com}
\and
\IEEEauthorblockN{Subhankar Mishra }
\IEEEauthorblockA{\textit{School of Computer Sciences} \\
\textit{National Institute of Science Education and Research}\\
Bhubaneswar, India \\
smishra@niser.ac.in}
}
\maketitle

\begin{abstract}
Forecast India’s economic growth has been traditionally an uncertain exercise. The indicators and factors affecting economic structures and the variables required to model that captures the situation correctly is point of concern. Although the forecast should be specific to the country we are looking at; however countries do have interlinkages among them. As the time series can be more volatile, and sometimes certain variables are unavailable; it is harder to predict for the developing economies as compared to stable and developed nations. However, it is very important to have accurate forecasts for economic growth for successful policy formations. One of the hypothesized indicators is the nighttime lights. Here, we aim to look for a relationship between GDP and Nighttime lights. Specifically we look at the DMSP and VIIRS dataset. We are finding relationship between various measures of economy.

\end{abstract}


\section{Introduction}
In ancient times, humans returned to their abodes as soon 
as the sun went down, wrapping up all activities they 
indulged in for livelihood. But with the advent of 
electricity, the scenario is changed. Human activities 
during the night which use light boost the economy of a 
nation. The status of a country’s economy is best described
by its Gross Domestic Product (GDP). It is important to look at status of economy using other data sources, because traditionally used data sources are not perfect. Sheer amount of transactions that happen makes it difficult to record and analyse it's contribution in economy. For instance, domestic workers hired in India are part of Informal Economy, which accounts for a huge section of employment in southeast Asia. Their contribution in economy is not counted in GDP. Hence official GDP measure might not be an actual representation of growth.\newline

There are three official methods to calculate a country's GDP called expenditure method, income method and production method.  Expenditure method arrives on GDP by calculating all the money spent by citizens in different sectors of economy which is called consumer spending. This method also takes government spending and net exports in account. Expenditure is done only if there is income, so income method could be applied to determine GDP. Production method measures the monetary value of all the goods and services produced in the country, which gives a distorted measure of GDP due to frequent price changes. But these methods don't account for off the books services, expenditure or income. Cash transactions in illegal trades like drug dealing which affect the economy but not considered in GDP. So we can say that, assumptions and approximation play a big role in calculating GDP.\newline

 Among many uses of the Nighttime lights, estimating the GDP is one of the ideas that has been explored. They could be used to assess traditional measures of GDP  as proved by Yingyao Hu and Jiaxiong Yao \cite{b1}. Jesús Crespo Cuaresma and others \cite{b2} have established the fact that Nighttime radiance could be used to estimate GDP even when official information is not available. Impact of major events like any natural disaster or war on economy could be measured by nighttime lights as done by Beyer and others \cite{b4}. Study done by Prakash, Shukla and others \cite{b7} proves seasonal variation of GDP and GSDP(Gross State Domestic Product) in India is well represented by nighttime lights.\newline

\begin{figure*}
        \begin{subfigure}[b]{0.5\textwidth}
                \includegraphics[width=\linewidth]{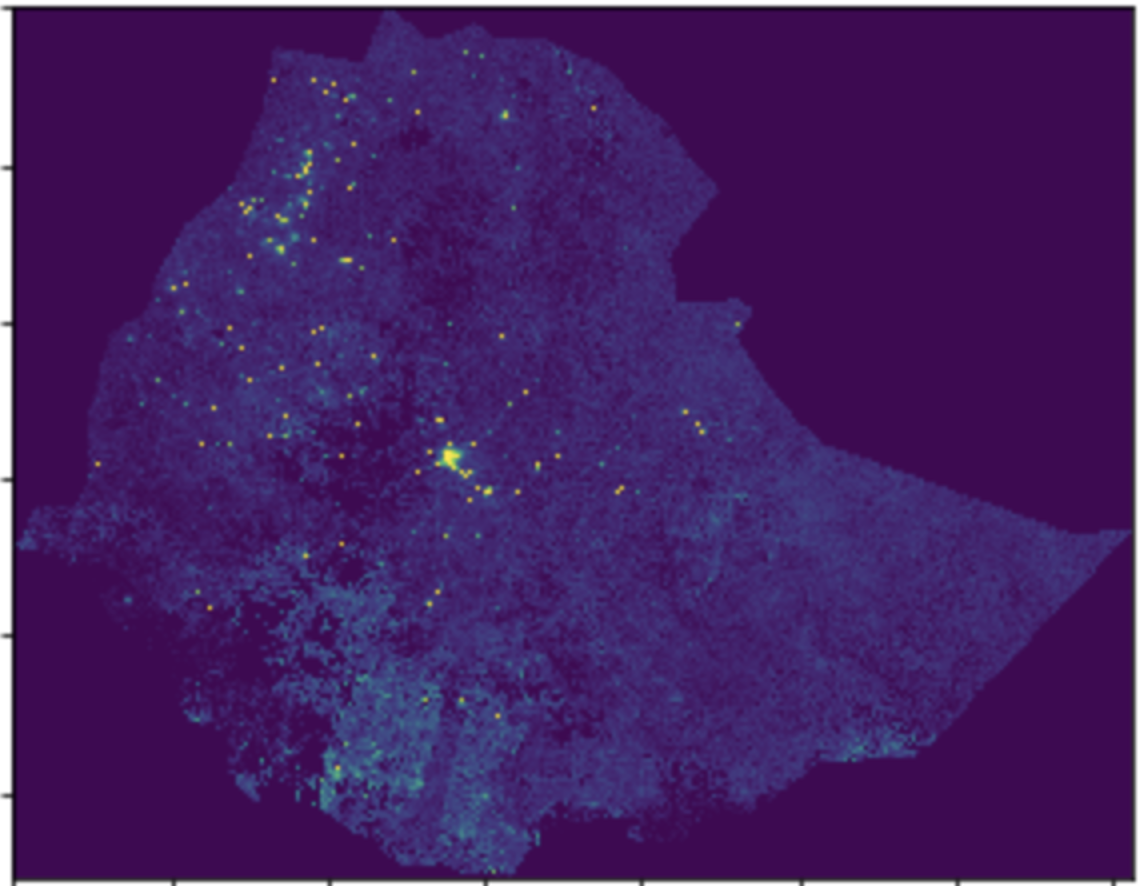}
                \caption{April 2012}
                \label{night:eth}
        \end{subfigure}%
        \begin{subfigure}[b]{0.5\textwidth}
                \includegraphics[width=\linewidth]{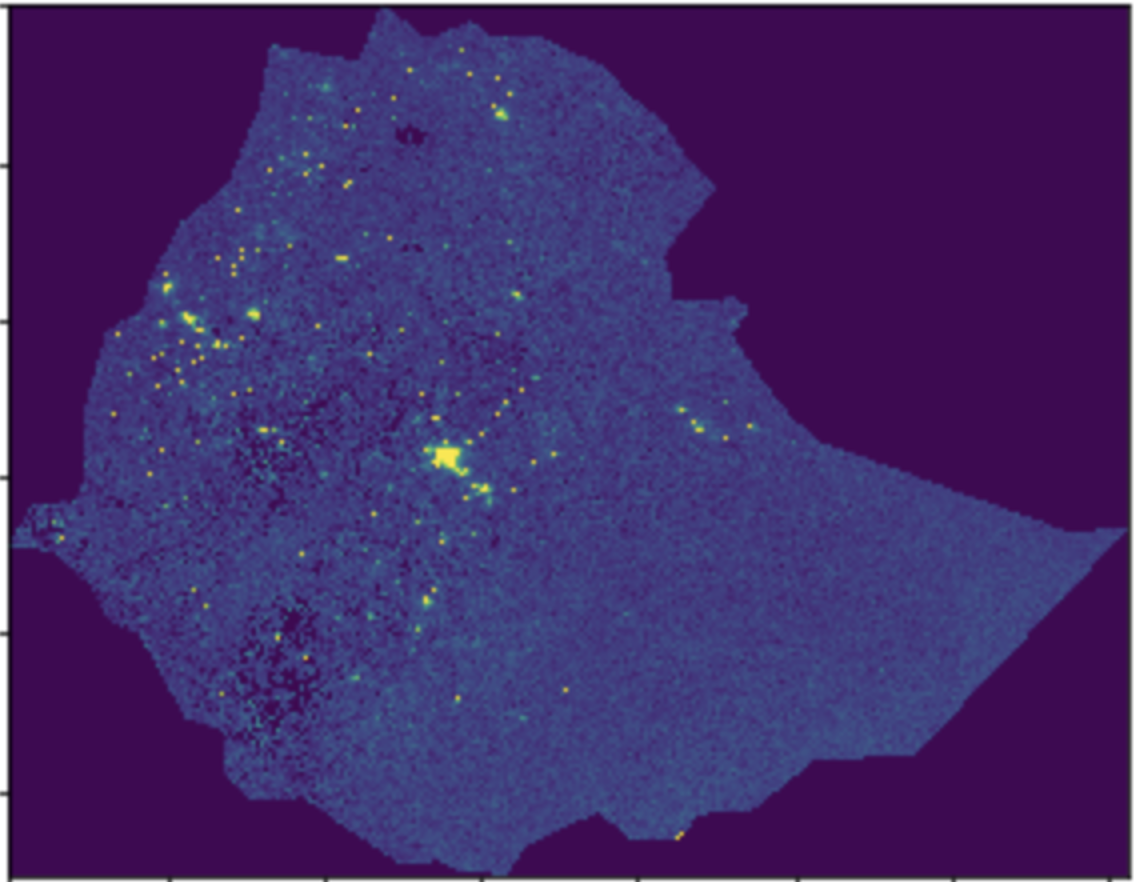}
                \caption{April 2019}
                \label{night:eth2}
        \end{subfigure}%
        \caption{Nighttime lights Of Ethiopia}\label{fig:night}
\end{figure*}

We try to find out up to what extent nighttime lights captured by different satellites in Earth’s orbit could serve as a proxy for economic data, by understanding trends between nightlights and GDP of few economies of the world (see figure \ref{fig:night}.
Our contribution is as follows:
\begin{itemize}
    \item Relationship between nighttime lights and the nominal GDP, real GDP and PPP.
    \item Developing a robust relationship by considering other economies close to India. 
\end{itemize}


The images captured by satellites show bright spaces at night, but they can be bright because of natural lights too. Gas flares, moonlight, Sun glare, are some examples. We need to first clean the data containing values from non-economic activities. Another issue is, while finding a relationship between nightlights and GDP of a region, we should consider the economic structure of the region. Economy of a particular region could be based on agriculture, which take place usually in the daylight, in such a case nightlights predictions will be wrong. Geographical and natural conditions also affect economy, those factors are not necessarily reflected by nightlights. 

\section{Related Works}

An IMF working paper \cite{b1} by Yingyao Hu and Jiaxiong Yao endeavors to mitigate errors in assessing GDP by proposing Nighttime Lights as new measure of GDP for middle-income countries. It proves Nighttime lights are a better measure for real GDP, by finding a non-linear relationship between GDP and Nighttime lights. Errors in measurement of real GDP were observed, which were not-surprisingly more for countries in turmoil, that led to overestimating or underestimating of GDP. To avoid such situations, official measures of GDP can be complemented by Nighttime lights. It takes several countries in consideration classifying them under high income, growing markets and those in turmoil due to various reasons like political instability, civil wars, terrorism,etc.and concludes high income countries like US have negligible errors, whereas  China, Brazil, Indonesia, and Pakistan have considerable errors.

Nighttime lights serve one more purpose of estimating GDP and various other economical aspects when reliable information is not available or no information is available. The work \cite{b2} of Jesús Crespo Cuaresma and others, of estimating poverty rates of North Korea whose reliable official statistics are not available. The relationship between Nightlights and GDP of southern parts of China was used to predict North Korea's GDP stating China's nightlights level is similar to that of North Korea. GDP was then estimated in all provinces of North Korea, which revealed the fact that it has lowest GDP in the world. Beta-Lorenz curve was used to predict income which in turn was used to find poverty rates of population. The highly heterogeneous distribution and fluctuation of nightlights over the years clearly suggest lack of development. Better picture could have been obtained by comparing nightlights intensity with a wider set of countries and using the closest one to estimate GDP.
\newline
The study done by Laveesh Bhandari and Koel Roychowdhary \cite{b3} finds out how economic activities in a developing country like INDIA affect relationship between nighttime lights and GDP. They studied the relationship on District level, classifying districts on type of their economy using DMSP-OLS images for nighttime lights data. They concluded Nighttime lights can indeed give an accurate idea of GDP even on a granular level. The errors encountered can be minimised by understanding sectoral GDP in depth and by using more efficient data on Nighttime lights, which we try to do by using VIIRS DNB Nighttime lights for year 2013 to 2019 along with DMSP-OLS for year 1992 to 2012.
\newline

This paper\cite{b4} proves strong correlation of Nighttime lights and GDP in South Asia and studies impact of major events in India, Nepal and Afghanistan on district level economy. Authors split GDP  into agricultural and non-agricultural GDP to predict GDP on basis of nighttime lights and population distribution, acknowledging the fact a large proportion of South Asia is agriculture based to predict GDP. Predicting GDP using monthly nighttime lights, they concluded economic blow suffered due to earthquakes in April and May 2015 in Nepal, was less severe than stated in official data and rural areas suffered worst. Also, districts affected due to disruption of trade with India experienced more decline of GDP growth compared to rest part of country. The effect of battle of Kunduz in Afghanistan on it's economy can be clearly seen by it's fluctuating nighttime lights intensity. GDP prediction on district level shows significant drop in quarterly GDP growth. The never ending debate of disadvantages of demonetization in India can be answered by nighttime lights, which show a slight trough in intensity graph for a short span of two months. But on district level GDP prediction says significant decrease is noticed for a short period.
\newline

There are many studies proving nighttime lights can be a proxy for economic indicators, but very few like \cite{b10} have tried to relate nighttime lights with sectoral economy. They have used NPP-VIIRS images, MODIS vegetation index and CORINE to find spatial distribution of agricultural and non-agricultural GDP of Turkey. Rasterized GDP, could be achieved by combining spatial data of all sectors in the country with nighttime lights. Similarly, the paper \cite{b7} found strong relation between nighttime lights and non-agricultural GDP.

\begin{table*}[htbp]
\centering
\caption{Summary of the related work}
\begin{tabular}{|p{0.3\textwidth}|p{0.15\textwidth}|p{0.15\textwidth}|p{0.3\textwidth}|}
\hline
Paper & Data Source  & Countries studied & Aim \\ \hline
Spatial distribution of GDP based on integrated NPP‑VIIRS
nighttime light and MODIS EVI data:a case study of Turkey & NPP-VIIRS  & Turkey & Rasterized GDP map of Turkey with Agricultural and Non-agricultural GDP   \\ \hline
What do we know about poverty in North Korea? & NPP VIIRS & North Korea & Estimate poverty rates of North Korea without reliable statistical economic data   \\ \hline
 Night Lights and Economic Activity in India: A study using
DMSP-OLS night time images & DMSP-OLS & India & Relation between economic activities in India at district level and nighttime lights   \\ \hline
Measuring Districts’ Monthly Economic
Activity from Outer Space & DMSP-OLS and NPP-VIIRS & Indian Subcontinent & Impact of Nepal earthquake, Afghanistan's war and Demonetization in India on Nighttime lights  \\ \hline
Night-time Luminosity: Does it Brighten
Understanding of Economic Activity in India? & NPP-VIIRS & India & Relation of nighttime lights with industrial production, State GDP and GDP growth of India    \\ \hline
 Illuminating Economic Growth & DMSP-OLS and NPP-VIIRS & Many countries  &  Proposes new measures to calculate GDP to mitigate errors in official GDP  \\ \hline
\textbf{Our work} & DMSP-OLS and NPP-VIIRS & All countries with GDP comparable with India & Relationship between Nighttime Lights and developing economies   \\ \hline
\end{tabular}
\end{table*}

\section{Background Theory}
 Real GDP, Nominal GDP, Purchasing Power Parity(PPP) ,GDP growth, and GDP per capita growth are official measures of nation's economy.
        \begin{enumerate}
        \item Nominal GDP is valuation of goods and services using current prices, also known as the current prices GDP.
        
        \item Real GDP is nominal GDP adjusted for inflation.
        \item  Purchasing power parity is used compare economies of different countries through a "basket of goods" approach.
        \item The GDP growth rate compares one year (or quarter) of a country's GDP to the previous year (or quarter) in order to measure growth of economy.
        
        \item Per capita gross domestic product is calculated by dividing the GDP of a country by its population. It is a global measure for measuring economic growth of countries.

        \end{enumerate}
\section{Data}
The intensity of Nighttime lights information is determined from images by Earth Observation Group, NOAA National Centers for Environmental Information (NCEI). Two programs of nighttime imagery were utilized which are known as Defense Meteorological Satellite Program Operational Linescan System (DMSP-OLS) and Visible Infrared Imaging Radiometer Suite (VIIRS) Day/Night Band (DNB). \newline
\subsection{DMSP}
    \begin{enumerate}
     \item DMSP information is gathered by the US Air Force Weather Agency. Image and data processing by NOAA's National Geophysical Data Center \cite{b5}. 
     \item Visible and infrared imagery from DMSP Operational Linescan System instruments swath an area of 3,000 km which pass between 19:30 to 20:30, catching zones with temperature between 190 to 310 Kelvins, with resolution of 0.55 km at high resolution mode and 2.7 km at low resolution mode.
     \item Cloud free composites in Geotiff format from year 1992 to 2013 of DMSP-OLS arrangement comprise of three kinds of images:
     \begin{enumerate}
      \item  Cloud-free coverages count the total number of observations. This image can be used to find regions with fewer observations.
       \item Average of visible band values with no further filtering.
       \item Average of only stable lights visible band values, which have zero background clamor, any precarious lights, for example, fires are eliminated. Average of stable lights visible band data values range from 1 - 63. These images are utilized for analysis because stable light sources like urban communities, towns, industrial regions, air terminals, stations, emergency clinics, educational institutions represent monetary exercises in evening time. 
       \end{enumerate}
       All these images are processed to remove all data affected by lighting from natural sources like sun glare, moonlight, cosmic light and observations with clouds.
   \end{enumerate}
\subsection{VIIRS}
\begin{enumerate}

  \item DMSP information was discontinued in 2013. From that point forward, Visible Infrared Imaging Radiometer Suite (VIIRS) Day/Night Band (DNB) which is operated by NASA-NOAA Joint Polar Satellite System (JPSS), set up in 2011 are utilized to compute evening lights. SNPP is the Suomi National Polar Partnership satellite flown by NASA and NOAA \cite{b6}. The essential imager on SNPP is the Visible Infrared Imaging Radiometer Suite (VIIRS).
  \item 
  VIIRS scans the Earth twice each day crossing the equator around 1:30 and 13:30. The 3,000 km swath width of the VIIRS instrument, allows for no gaps in coverage. The VIIRS instrument works at two spatial resolutions 500 m and 1 Km.
 \item 
  The monthly composites from year 2013 to 2019 are used, which comprise of  6 geotiff tiles for each month. Annual composites are accessible for only 2014 and 2015. The tiles are 15 arc-second geographic matrices in geotiff format . The tiles are cut at the equator and each span 120 degrees of latitude. Each tile is a set of images containing average DNB radiance values with units in $nanoWatts/cm^2/sr$, total number of cloud-free observations used in the average, and the number of total DNB observations regardless of cloud cover.
  \item In the monthly composites, average radiance of numerous parts of globe is observed to be null or higher than expected because of natural factors like cloud-cover, especially in the tropical regions, or due to sunlight,  as occurs toward the poles in their mid year months. To avoid these shortcomings, we have used VIIRS images only of September, October and November for all years.
 \end{enumerate}

\begin{figure}[htbp]
\centerline{\includegraphics[width=\linewidth]{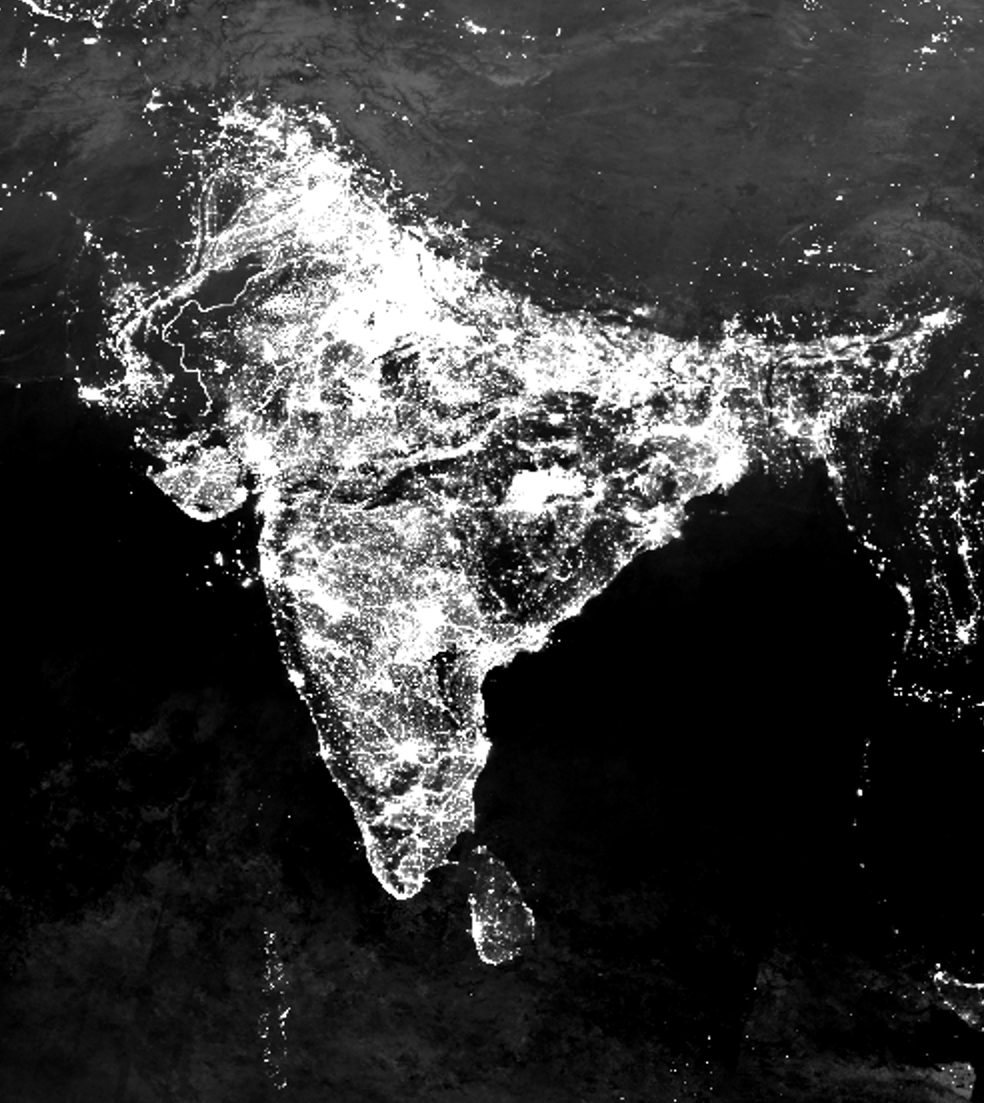}}
\caption{VIIRS data for for Tile 3 for the month of December, 2015. (cropped for India)}
\label{fig1}
\end{figure}

A sample VIIRS image is shown in fig \ref{fig1}.
 
\begin{table}[htbp]
\centering
\caption{Difference Between DMSP-OLS And SNPP VIIRS}
\begin{tabular}{|l|l|l|}
\hline
\textbf{Factors}      & \textbf{DMSP-OLS} & \textbf{SNPP VIIRS}                       \\ \hline
Operators & US Military & NASA-NOAA \\ \hline
Image Format & Geotiff  & 6 Geotiff  tiles                   \\ \hline
Units        & 0-63     & nanoWatts/cm\textasciicircum{}2/sr \\ \hline
 Spatial resolution  &  0.55 km and 2.7 Km      &   0.5 Km and 1 Km                                \\ \hline
 Active years   &   1992-2013      &   2012-till today                                 \\ \hline
  Overpass time           &  19:30 to 20:30        &  1:30                                  \\ \hline
\end{tabular}
\end{table}
\subsection{Statistical Data}
GDP\cite{b8} and population data for countries\cite{b9} is taken from World Bank.

\section{Methodology}

\begin{enumerate}
    \item Firstly we identified countries of which we wish to find relation between economy of a country and nighttime lights. Real GDP, Nominal GDP, Purchasing Power Parity(PPP) ,GDP growth, and GDP per capita growth are the variables we wish to relate with nighttime lights. Utilizing World Bank's data of GDP and GDP growth, we selected around 60 countries having GDP values around that of India's values of the year 2018.\newline

 \item  Source of our nighttime light intensity is DMSP and VIIRS, so we downloaded all the DMSP and VIIRS images.
 
 \item It was observed upon analysis that monthly VIIRS images had flaws such as some countries or part of the country was missing for summer months for countries closer to north pole, or image was too bright in January and February due to reflection from snow. Also, literature survey proved us right. This led to rejection of inaccurate monthly images and average of nighttime light intensity of September, October and November was calculated for each year.
 
 \item
 
  Satellite imagery has data in raster format stored in GeoTIFF images. Geographic information systems use GeoTIFF image format which has georeferencing information embedded within the image file. Raster consists of  pixels organized into a grid where each pixel contains radiance value of a particular area on ground. Area of the pixel varies according to spatial resolution of satellite.
  \newline 
 Rasterio library is used to access geospatial raster datasets by using GeoJson files. We downloaded GeoJson files for selected countries and used those as input for reading raster as 2-d numpy array. Value of each pixel is an element of the array, all elements were added to calculate nighttime lights intensity for each country and all months from year 1992 to 2018. average of nighttime light intensity of September, October and November was calculated for each year.

\item Data from DMSP and VIIRS has certain differences for example unit of the former is digital band number and that of the latter is nanoWatts/(cm2 sr) and spatial resolution of VIIRS is finer than DMSP. We used an algorithm to coordinate DMSP data for years 1992 to 2012 with VIIRS data for years 2013-2018. We subtracted average of difference between VIIRS and DMSP nighttime light intensity from the common years 2012 and 2013 from nighttime light intensity of subsequent years.
\item Economy of nation depends on its human resources along with nighttime lights, so we downloaded population of selected countries from world bank development indicators and use it as independent variable for analysis.
\end{enumerate}

\begin{figure}[htbp]
\centerline{\includegraphics[width=0.5\linewidth]{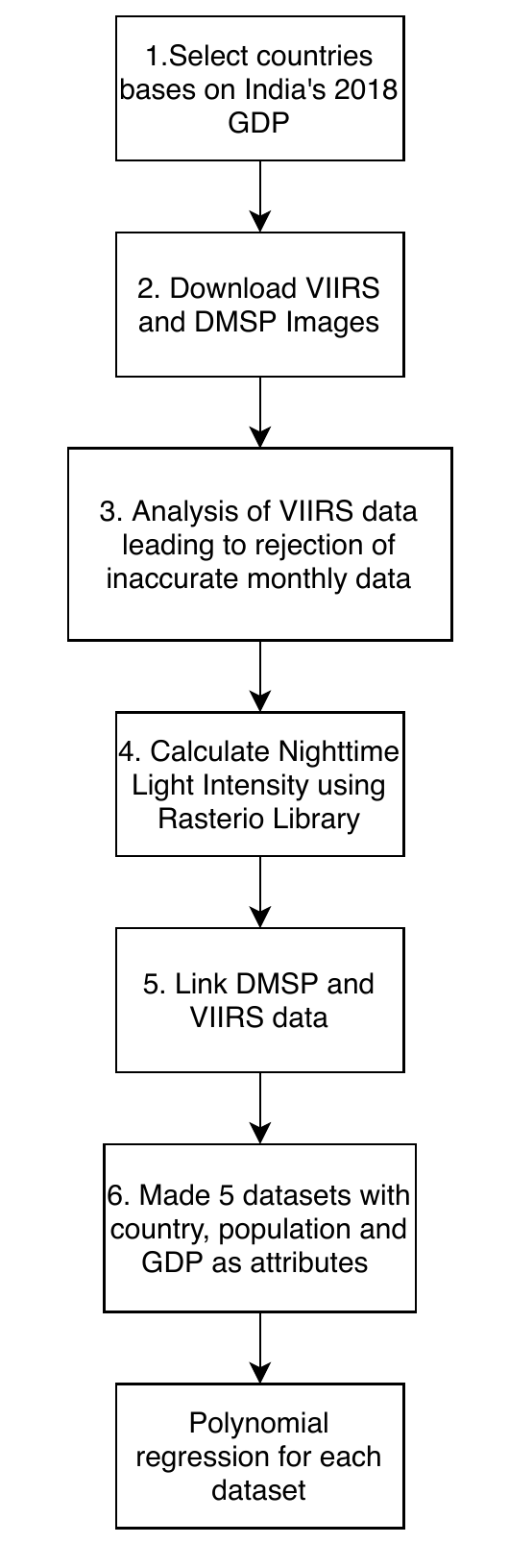}}
\caption{Flowchart}
\label{fig2}
\end{figure}

\begin{table*}[ht]
\centering
\caption{ Result of polynomial regression with Luminosity and Population }
\label{tab:Table 2}
\begin{tabular}{|l|l|l|l|l|l|l|l|}
\hline
  Y              & \multicolumn{5}{l|}{Coefficients}& Intercept & $R^2$\\ \hline
         & $x_{1}$   & $x_{2}$   & $x_{1}^2$   & $x_{1}x_{2}$   &  $x_{2}^2$ &   & \\ \hline
Real GDP      & 0.432       &  0.411    & -0.304     &  3.024    & -0.941   &  0.025  &  0.625  \\ \hline
Nominal GDP   &  0.359    & 0.277     & -0.439       & 3.084    & -0.802   & 0.021   & 0.543     \\ \hline
PPP GDP               &  0.310    &  -0.017   &  -0.394    &  3.150  & -0.356   & 0.04  & 0.699 \\ \hline
GDP Growth            &  -0.264    &  0.312    &  0.752  & -0.609  & -0.0.19   & 0.790  & 0.022 \\ \hline
Per capita GDP growth &  -0.050     &  0.012    &  0.255    & -0.247   & 0.094   &  0.370  & 0.007 \\ \hline
\end{tabular}
\end{table*}

\begin{table*}[ht]
\centering
\caption{ Result of polynomial regression with Luminosity and Year }
\label{tab:Table 3}
\begin{tabular}{|l|l|l|l|l|l|l|l|}
\hline
  Y              & \multicolumn{5}{l|}{Coefficients}& Intercept & $R^2$\\ \hline
         & $x_{1}$   & $x_{3}$   & $x_{1}^2$   & $x_{1}x_{3}$   &  $x_{3}^2$ &   & \\ \hline
Real GDP      & 0.104       &  -0.163    & 0.317     &  1.269    & 0.151   &  0.057  &  0.672  \\ \hline
Nominal GDP   &  -0.160    & -0.156     & 0.280       & 1.532    & 0.146   & 0.049   & 0.681     \\ \hline
PPP GDP               &  0.219    &  -0.141   &  -0.141    &  1.440  & 0.177   & 0.025  & 0.814 \\ \hline
GDP Growth            &  0.129    &  0.112    &  0.152  & -0.255  & -0.055   & 0.763  & 0.099 \\ \hline
Per capita GDP growth &  0.018     &  0.133    &  0.058    & -0.049   & -0.111   &  0.341  & 0.105 \\ \hline
\end{tabular}
\end{table*}

\section{Analysis}
Multivariate polynomial regression model in two degree was used for analysis. This model is used when scatter plot of output variable is of non-linear nature. Here output variables are GDP measures which have curvilinear plots. Mathematical representation of this model is:
\newline
\begin{equation}
y = a_{1}x_{1} + a_{2}x_{2} + a_{3}x_{1}x_{2} + a_{4}x_{1}^2 + a_{5}x_{2}^2
\label{eq1}
\end{equation}
where y is dependent variable and $x_{1}$ and $x_{2}$ are independent variables or determinants. Equation \eqref{eq1} could be re-written as:
\begin{equation}
y = a_{1}z_{1} + a_{2}z_{2} + a_{3}z_{3} + a_{4}z_{4} + a_{5}z_{5}
\label{eq2}
\end{equation}
where $z_{1}$=$x_{1}$, $z_{2}$=$x_{2}$, $z_{3}$= $x_{1}$*$x_{2}$, $z_{4}$=$x_{1}^2$, $z_{5}$=$x_{2}^2$
which is a multivariate linear regression model whose coefficients are estimated using least square method. We used scikit-learn python module to perform analysis. \newline

Polynomial regression was performed on five datasets. Each dataset has four attributes of which country, year and population are common. Fourth attribute of dataset are Real GDP , nominal GDP, PPP, GDP growth and, per capita GDP growth which are also dependent variable of the polynomials. Table.~\ref{tab:Table 2} shows result of two degree polynomial regression with population and luminosity as independent variable and table .~\ref{tab:Table 3} shows that with year and luminosity as independent variable, where $x_1$ denotes luminosity, $x_2$ denotes population and $x_3$ denotes year of the respective country and year. R-square score depicts goodness of fit of the fitted regression line, also known as coefficient of determination.  Higher R-square shows better fitted data. \newline
It is observed that R square is more for GDP measures in case of year as an independent variable. Quantitatively, R square of nominal GDP is 0.681 with year and luminosity which is statistically significant than R square of 0.543 with population and luminosity. Real GDP correlates well with nighttime lights in both cases. PPP GDP, has highest R square 0.814 with year. GDP growth and GDP per capita growth have very low R-square indicates change in GDP growth is not represented well by change in nighttime lights.

\section{Conclusion}
This paper shows that luminosity obtained by DMSP and VIIRS images have a strong relationship with official GDP measures of economies ranking around India. 
Regression analysis establishes the fact nighttime lights and year exhibit a strong capability to model economic indicators. This shows that population solely can't affect the economic activity of any country, their are other factors at play that change with time and affect economy of all countries. PPP, which is a suitable measure to compare economies has highest R square with year and luminosity as independent variables which implies economies ranking around India have strong correlation between GDP and nighttime lights. We believe our study would be useful to devise economic policies for developing economies as the relationship found could be used to forecast economic indicators.
However, there are some limitations of nighttime lights for example the type of economy plays a role here, a tourism based or industrial economy will have higher radiance at night as compared to agriculture based economy. So the polynomial regression model used for analysis follows a basic approach and requires few determinants, better results are possible with inclusion of more factors.  

\bibliography{ref.bib} 

\begin{thebibliography}{10}

\bibitem{b1}
Y.~Hu and J.~Yao, {\em Illuminating economic growth}.
\newblock International Monetary Fund, 2019.

\bibitem{b2}
J.~C. Cuaresma, O.~Danylo, S.~Fritz, M.~Hofer, H.~Kharas, and J.~C.~L. Bayas,
  ``What do we know about poverty in north korea?,'' {\em Palgrave
  Communications}, vol.~6, no.~1, pp.~1--8, 2020.

\bibitem{b4}
R.~C. Beyer, E.~Chhabra, V.~Galdo, and M.~Rama, ``Measuring districts' monthly
  economic activity from outer space,'' {\em World Bank}, 2018.

\bibitem{b7}
A.~Prakash, A.~K. Shukla, C.~Bhowmick, and R.~C.~M. Beyer, ``Night-time
  luminosity: Does it brighten understanding of economic activity in india?,''
  {\em Reserve Bank of India occasional Papers}, vol.~40, p.~1, 2019.

\bibitem{b3}
L.~Bhandari and K.~Roychowdhury, ``Night lights and economic activity in india:
  A study using dmsp-ols night time images,'' {\em Proceedings of the
  Asia-Pacific advanced network}, vol.~32, no.~0, p.~218, 2011.

\bibitem{b10}
E.~Ustaoglu, R.~Bovk{\i}r, and A.~Ayd{\i}noglu, ``Spatial distribution of gdp
  based on integrated nps-viirs nighttime light and modis evi data: a case
  study of turkey,'' {\em Environment, Development and Sustainability},
  pp.~1--35, 2020.

\bibitem{b5}
``Version 4 dmsp-ols nighttime lights time series.''
  \url{https://ngdc.noaa.gov/eog/dmsp/downloadV4composites.html}, 1992.
\newblock Accessed on: 2020-06-27.

\bibitem{b6}
``Version 1 viirs day/night band nighttime lights,'' 2012.

\bibitem{b8}
``World bank.'' \url{https://data.worldbank.org/indicator/NY.GDP.MKTP.PP.CD}.
\newblock Accessed on: 2020-11-30.

\bibitem{b9}
``World development indicators database, world bank.''
  \url{https://databank.worldbank.org/source/world-development-indicators}.
\newblock Accessed on: 2020-11-30.

\end{thebibliography}
\bibliographystyle{ieeetr}



  


\end{document}